\documentclass[
twocolumn,
aps,prd,
nofootinbib,
superscriptaddress,
showpacs,
tightenlines,
]{revtex4}
\usepackage{hyperref}
\usepackage{amsmath}
\usepackage{amssymb}
\usepackage{bm}
\usepackage{color,graphicx}
\usepackage{slashed}
\usepackage{caption}
\usepackage{float}
\usepackage{subcaption}
\usepackage[compat=1.1.0]{tikz-feynman}

\begin{document}

\title{Connected and disconnected contributions to nucleon form factors and parton distributions}

\author{Zaki Panjsheeri} 
\email{zap2nd@virginia.edu}
\affiliation{Department of Physics, University of Virginia, Charlottesville, VA 22904, USA.}

\author{Saraswati Pandey} 
\affiliation{Department of Physics, University of Virginia, Charlottesville, VA 22904, USA.}

\author{Brannon Semp} 
\affiliation{Department of Physics, University of Virginia, Charlottesville, VA 22904, USA.}

\author{Simonetta Liuti} 
\email{sl4y@virginia.edu}
\affiliation{Department of Physics, University of Virginia, Charlottesville, VA 22904, USA.}


\begin{abstract}
Using the framework of generalized parton distribution, we provide a unified interpretation of the connected and disconnected contributions from the ab-initio Euclidean
path-integral formulation of the hadronic tensor in both the nucleon elastic form factors and the parton distribution functions. We develop a phenomenology to elucidate non-perturbative contributions to deep inelastic structure functions, which can be extended to observables in heavy-ion collisions probing baryon junctions.   
\end{abstract}

\maketitle

The nucleon electromagnetic form factors (FFs) observed in elastic electron-nucleon scattering experiments 
have been the first observables to reveal a non-trivial structure of charge distributions,
leading to decades of modeling and physical interpretations of the relative configurations of the positive and negative charges. 
%
Notwithstanding, perhaps surprisingly, several fundamental questions still remain to be answered concerning the proton and neutron interior's spatial flavor structure, which is in principle accessible through Fourier transformation of the FFs (regarding the connection through Fourier transformation between the form factor and the spatial density we refer the reader to the discussion in Refs.\cite{Jaffe:2020ebz,Panteleeva:2022khw}).
In a QCD-based description, it has been assumed that the nucleon FFs are dominated by up and down valence quarks, since the presence of the strange and charm quark flavors depends on the asymmetry of their $q$ and $\bar{q}$ distributions, which is deemed to be small from the experimental evidence collected  so far \cite{Maas:2017snj}. 
Under this assumption, the $u$/$d$ flavor separated content of the nucleon FFs was provided for the first time in Ref.\cite{Cates:2011pz} and confirmed in Refs.\cite{Arrington:2011kb,Qattan:2024pco}, revealing unexpected behaviors of the $u$ and $d$ quarks at momentum transfer higher than $\approx 1$ GeV$^2$, for both the Dirac and Pauli form factors and their weighted ratios. 
In the antiquark sector, $\bar{u}$/$\bar{d}$ flavor asymmetry in the proton was observed through the violation of the Gottfried sum rule (\cite{SeaQuest:2021zxb,Geesaman:2018ixo} and references therein) at the short-distance scale accessed in deep inelastic scattering, where it cannot be ascribed to chiral effective field theory mechanisms.
Finally, the measurement of a 10$\%$ excess of baryons over antibaryons in heavy ion collisions 
highlights the role of baryon junction dynamics  in baryon number transport, and brings to the forefront fundamental questions on the dynamics underlying the spatial configurations of quarks and gluons inside the proton \cite{Magdy:2025udq}.
%

These complementary insights into the proton's internal structure prompt us to ask whether they are manifestations of the same underlying dynamics. It also motivates the exploration of how lattice QCD (LQCD) can be most efficiently employed to disentangle the effects of individual quark, antiquark and gluon contributions to nucleon structure. In the Euclidean path-integral formulation of the hadronic tensor in QCD
we distinguish two kinds of sea parton contributions from connected and disconnected insertions, respectively \cite{Liu:2012ch,Liu:1999ak, Leinweber:2022guz,Geesaman:2018ixo}. 
The latter have been evaluated most recently for the elastic form factors in Refs.\cite{Alexandrou:2025vto,Syritsyn:2025fiu}. Their possible impact on PDF analyses was estimated in the phenomenological study of Ref.\cite{Hou:2019efy} where it was inferred that the disconnected sea is responsible for most of the PDFs behavior at the lowest $x$ values. 
Here we  revisit the extraction of connected and disconnected distributions from experimental data. Our approach is based on the observation that the information on the proton's deep internal structure obtained from either FFs or PDFs, has a common origin in QCD correlations known as generalized parton distributions (GPDs). 
GPDs provide a unified description of hadron structure, encompassing the information contained separately in FFs and PDFs.
Based on a GPD model constrained by experimental data and LQCD \cite{Panjsheeri:2025vpa}, we provide a simultaneous description of the disconnected contribution to both the FFs and the PDFs. 
%



GPDs parameterize nonlocal, gauge-invariant quark and gluon correlation functions in the nucleon. In particular, they are defined from off-forward (nonzero momentum transfer) matrix elements of bilocal light-cone operators between proton states with definite helicity. Here we focus on the quark correlator for the vector operator, $\gamma^+$, which is parametrized in terms of GPDs $H_q$ (helicity conserving), and $E_q$ (helicity flip) (we refer the reader to Refs.\cite{Diehl:2003ny,Belitsky:2005qn,Kumericki:2016ehc} for reviews). At any given QCD scale, $Q^2$, GPDs depend on: $x$ – the average longitudinal momentum fraction of the parton, $\xi$  – the skewness (fraction of the longitudinal momentum transfer)
and $t$ – momentum-transfer squared to the hadron.
In the forward limit, $\xi,t=0$, the following correspondence holds between the unpolarized GPD $H_q$,  and the unpolarized PDFs $q(x)$,
\begin{eqnarray}
    H_q(x,0,0) = q(x)
    \label{eq:forward}
\end{eqnarray}
The elastic form factors correspond to the $n=1$ Mellin moments,
\begin{subequations}
\begin{eqnarray}
\label{eq:dirac}
F_1^q(t) = A_{10}^q(t) = \int_{-1}^{1} dx H_q(x,\xi,t) , \\
\label{eq:pauli}
F_2^q(t) = B_{10}^q(t) = \int_{-1}^{1} dx E_q(x,\xi,t) 
\end{eqnarray}
\end{subequations}
while for $n=2$, 
\begin{subequations}
\begin{eqnarray}
\label{eq:A20}
M_2^{q,H}(\xi,t) =  A_{20}^q(t) + (2\xi)^2 C_{20}^q(t) = \int_{-1}^{1} dx \, x \,H_q  
\\
\label{eq:B20}
M_2^{q,E}(\xi,t) =   B_{20}^q(t) - (2\xi)^2 C_{20}^q(t)  =  \int_{-1}^{1} dx \, x \, E_q 
\end{eqnarray}
\end{subequations}
The proton GPD (including the PDF in the forward limit) is written  as,
\begin{eqnarray}
\label{eq:Hp}
H_p  =  \sum_q e_q^2 \, (H_q+H_{\bar{q}}) 
\end{eqnarray}
where $e_q$ is the quark charge, and we consider contributions up to the strange quark. 
%
Crossing symmetry relations with respect to $x \rightarrow -x$, allow us to introduce the flavor non-singlet, and flavor singlet distributions  \cite{GolecBiernat:1998ja,Goldstein:2010gu,Kriesten:2019jep}. 
We define the unpolarized anti-quark GPD, $H_{\bar{q}}(x,\xi,t)$ as, 
\begin{eqnarray}
    H_{\bar{q}}(x,\xi,t) = - H_{{q}}(-x,\xi,t) 
\end{eqnarray}
The $``+"$ and $``-"$ contributions are defined as, 
\begin{subequations}
\label{eq:Hplusminus}
\begin{eqnarray}
    H^+_q(x,\xi,t) = H_{{q}}(x,\xi,t)  + H_{\bar{q}}(x,\xi,t) \\
    H^-_q(x,\xi,t) = H_{{q}}(x,\xi,t)  - H_{\bar{q}}(x,\xi,t)
\end{eqnarray}
\end{subequations}
with symmetry properties for $x \rightarrow -x$ (we omit writing the dependence on the $\xi,t$ variables, which are not symmetry arguments),
\begin{subequations}
\label{eq:Hplusminu}
\begin{eqnarray}
H^-_q(-x) &= & H^-_q(x) \quad\quad symmetric \; \\
H^+_q(-x) &=& -H^+_q(x) \quad antisymmetric 
 \end{eqnarray}
\end{subequations}
In the Kuti-Weisskopf ansatz, $H_{\bar{q}} = H_{q_{sea}}$ \cite{Kuti:1971ph}, 
\begin{subequations}
\label{eq:Kuti}
\begin{eqnarray}
H^-_{q}(x)  &=&   H_{q_v}(x) \\
H^+_{q}(x) &= & H_{q_v}(x) + 2 H_{q_{sea}}(x) .
\end{eqnarray}
\end{subequations}
Therefore, the FF \eqref{eq:dirac}, is given solely by the $``-"$ (symmetric) component, being transparent to the sea quark contribution; the $n=2$ moment \eqref{eq:A20}, is given by the $``+"$ (antisymmetric) 
sea  component.
LQCD evaluations can, however, differ from the Kuti-Weisskopf ansatz,  depending on how the sea quarks are introduced, namely, the contribution of the connected sea quarks (CS) to the PDFs/GPDs is defined as a four point current-current
correlator for the nucleon through a ``connected" insertion of the current on a valence quark line. The ``disconnected" sea quarks (DS), by contrast, originate in a quark loop from vacuum polarization,  
thus introducing an additional degree of freedom to be considered in the evaluation of Eqs.\eqref{eq:Hplusminu} \cite{Liu:2012ch, Liu:1999ak}. 
A non-negligible ``symmetric" DS  appears in the LQCD form factor results from Ref. \cite{Alexandrou:2025vto}, with 
the DS becoming relatively more important compared to the CS at large four-momentum transfer, $|t|$.    
Analyzing these contributions in a GPD-based underlying picture, we replace Eq.\eqref{eq:dirac} for the $q=u+d$ component with,
\begin{eqnarray}
\label{eq:F1_CSDS}
    F_1^{u+d} &=& [F_1^{u+d}]^C + [F_1^{u+d}]^D =  \int_{-1}^{1} dx \, \left[H^C_{u+d} + H^D_{u+d} \right] \nonumber \\
        &&\equiv \int_0^1 dx \, H_{u_{v}+d_{v}} + \int_{0}^1 dx \, H^{D,{sym}}_{u+d} 
\end{eqnarray}
where we define, 
\begin{eqnarray}
    H_{u+d}^C =  H_{u_{v}+d_{v}}  + H_{u_{sea}+d_{sea}}^C ,
\end{eqnarray}
observing the symmetries in Eqs.\eqref{eq:Kuti}, and we introduce 
$H^{D,{sym}}_{u+d}$, a symmetric with respect to $x=0$, component of DS.
Similarly, we obtain for $M_2^{q,H}$,
\begin{eqnarray}
\label{eq:M2_CSDS}
   && M_2^{u+d,H} =  \int_{-1}^{1} dx \, x \left[H^C_{u+d} + H^D_{u+d} \right] \nonumber 
    \\ &  \equiv &   \int_{0}^1 dx \, x \left[H_{u+d}^{C,+} \right] + \int_{0}^1 dx \, x \left[ H^{D,{antisym}}_{u+d} \right],
\end{eqnarray}
The contribution of the DS two additional components, $H_{u+d}^{D,sym}$, Eq.\eqref{eq:F1_CSDS} and $H_{u+d}^{D,antisym}$,  Eq.\eqref{eq:M2_CSDS} 
can be gauged using a suitable parametrization constrained by data and LQCD results, as a guideline. 
Using the reggeized-diquark model, we obtained the following parametric forms of the GPDs $H_q^-$ and $E_q^-$, at an initial QCD scale $Q_o^2 = 0.58 $ GeV$^2$,
\begin{subequations}
\label{eq:GPD_minus}
\begin{eqnarray}
    H_{q}^- & = & {\cal N}^H_q  x^{-\big[\alpha_q + \alpha_q'^H \, t \, (1-x)^{p_q^H} \big] } F_q^H(x,\xi,t) \\
    E_{q}^- & = & {\cal N}^E_q  x^{-\big[\alpha_q + \alpha_q'^E \, t \, (1-x)^{p_q^E} \big] } \, F_q^E(x,\xi,t) 
\end{eqnarray}
\end{subequations}
for $q=u,d$; $F_q^{H,E}(x,\xi,t)$ is a modulating form depending on the mass parameters, $m_q$, $M_\Lambda^q$, $M_X^q$.  Their full expressions, including the forward limit, are given in Ref.\cite{Panjsheeri:2025vpa}.
\footnote{Perturbative-QCD evolved grids in $X,\zeta,t$, to $Q^2>Q_o^2$ can be found in \noindent\textbf{\textcolor{black}{\url{https://github.com/Exclaim-Collaboration/UVA2} }}}
The parameters were determined from a global fit using the standard curve fitting procedure of MIGRAD – Minuit’s principal optimization algorithm to efficiently minimize the chi-square through the inverse Hessian matrix \cite{Panjsheeri:2025vpa,iminuit}.
For the symmetric DS component we take the form,  
\begin{eqnarray}
\label{eq:minus_D_model}
 (GPD)^{D, sym}_q ={\cal N}_D^{-}   \, x^{-(\alpha_{\bar{q}} + \alpha't)} ,  
\end{eqnarray}
that can be directly be compared to the $D,sym$ contributions to the form factors given by the integrals in Eqs.\eqref{eq:F1_CSDS} at the higher scale of $Q^2=4$ GeV$^2$ for $(GPD)=H,E$. Note that, despite the simplicity of the Regge-phenomenology–based ansatz, we explicitly incorporate correlated $x$- and $t$-dependence.
\begin{table}[H]
    \centering
    \begin{tabular}{lcccc}
\hline
\hline
    {\bf GPD} \quad \quad & $\alpha$ \quad & $\alpha^{\prime}$ \quad & $p$ & $\mathcal{N}$ \quad   \\
    \hline 
    \hline
      $H_{u}^-$ &  0.427 & 0.890(8) & 0.950(27) & 5.93  \\
      $H_{d}^-$ &   0.33 & 0.869(33) & 0.15(14) & 2.2  \\
      $H_{u}^{D, sym}$ &  1.153 &  $0.080(18)$ & $-$ & $0.629(33) \times 10^{-3}$  \\
      $H_{d}^{D, sym}$ &  1.150 & $0.080(18)$ & $-$ & $0.629(33) \times 10 ^{-3}$ \\
      \hline
      $H_{\bar{u}}^C$ &  1.153 & $0.35(9)^\star$ & $0.25(18) \times 10$ $^\star$ & 0.19  \\
      $H_{\bar{d}}^C$ &   1.150 & $0.14(11)^\star$ & 2(1)$^\star$ & 1.1  \\
      $H_{u}^{D, antisym}$ &  1.170 &  $0.080(18)$ & $-$ & $0.0453(83)$  \\
      $H_{d}^{D, antisym}$ &  1.168 & $0.080(18)$ & $-$ & $0.0453(83)$ \\
           \hline
      $E_{u}^-$  & 0.427 & 1.62(11) & 1.76(9) & 6.87(32) \\
      $E_{d}^-$  & 0.33& 0.540(9) & 0.10(23) & $-3.55(5)$ \\
      $E_{u}^{D, sym}$ &  1.153 &  $0.427(23) $ & $-$ & $-4.62(19)\times 10^{-3}$  \\
      $E_{d}^{D, sym}$ &  1.150 & $0.427(23)$ & $-$ & $-4.62(19)\times 10^{-3}$ \\
      \hline
      $E_{\bar{u}}^C$  & 0.427 & 0.00(19)$^\star$ & 4(4)$^\star$ & $-0.314(6)$ \\
      $E_{\bar{d}}^C$  & 0.33 & 0.00(6)$^\star$ & 5(5)$^\star$ & $-0.46(20)$ \\
      $E_{u}^{D, antisym}$ &  1.170 &  $0.427(33)$ & $-$ & $-0.016(16)$  \\
      $E_{d}^{D, antisym}$ &  1.168 & $0.427(33)$ & $-$ & $-0.016(16)$ \\
      \hline
      \hline
          \end{tabular}
    \caption{Fitted values of the parameters for the connected and disconnected contributions at the initial UVA2 parametrization scale, $Q_{0}^{2} = 0.58 \: \mathrm{GeV}^{2}$ \cite{Panjsheeri:2025vpa} for the $``-"$ and the connected $\bar{u}$ and $\bar{d}$ components; the $D, sym/antisym$ components were fitted at the LQCD scale of $Q^2=4$ GeV$^2$. Parameters with the $\star$ only to apply to the $t$ dependence. The errors on the parameters for the $``-"$ and the connected $\bar{u}, \bar{d}$ components were evaluated in Ref.\cite{Panjsheeri:2025vpa}; the error on the $D, sym/antisym$ components are from the fit and normalization to the LQCD values of the form factors \cite{Alexandrou:2020sml,Alexandrou:2025vto}.}
    \label{tab:parameters}
\end{table}
\begin{figure}[h!]
    \includegraphics[width=0.7\linewidth]{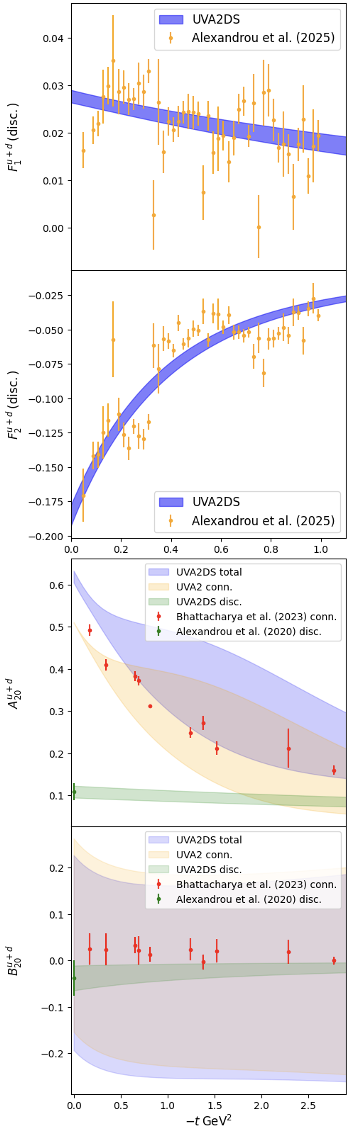}
\caption{Fit of the GPD models for the symmetric, Eq.\eqref{eq:minus_D_model}, distributions to the disconnected insertion component of the Dirac and Pauli form factors from Ref. \cite{Alexandrou:2025vto} (upper panels) and the antisymmetric, Eq. \eqref{eq:plus_D_model}, distributions to the disconnected insertion component of the $A_{20}, B_{20}$ moments Ref.\cite{Alexandrou:2020sml} (lower panels).}
    \label{fig:formf}
\end{figure}
\begin{figure}[h!]
    \includegraphics[width=0.7\linewidth]{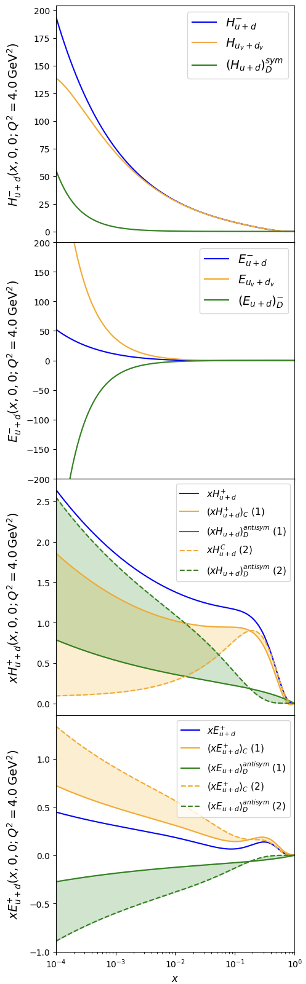}
\caption{Momentum fraction $x$ dependence predictions for the $+$ and $-$ components of the GPDs $H, E$. The blue lines give the total distributions, while the orange and green lines are the connected and disconnected contributions, respectively, obtained in this work.}
    \label{fig:PDF}
\end{figure}
Similarly, the antisymmetric component can be obtained  by fitting 
the second moment of $H_{q}^+$, $M_2^{q,H}(0,t) \equiv A_{20}^q(t)$, Eq.\eqref{eq:A20}. However, in this case, LQCD results for the separated $CS$ and $DS$ insertions are only available at $t=0$,  for the $u,d,s$ and $c$ components \cite{Alexandrou:2020sml}. 

%
%
%
Using our GPD model we define the CS GPD at the initial scale $Q_o^2 = 0.58$ GeV$^2$, as,
\begin{eqnarray}
(GPD)^C  =  {\cal N}^H_{\bar{q}}  x^{-\big[\alpha_{\bar{q}} + \alpha_{\bar{q}}'^H \, t \, (1-x)^{p_{\bar{q}}^H} \big] } F_{\bar{q}}^H(x,\xi,t) , 
\end{eqnarray}
and we evolve it to the $Q^2=4$ GeV$^2$ the LQCD results.
For the DS contributions, we adopt the following Regge-based form directly at $Q^2=4$ GeV$^2$, 
%
\begin{eqnarray}
\label{eq:plus_D_model}
 (GPD)^{D,antisym}  =  {\cal N}_D^{H,antisym} \, x^{-(\alpha_{\bar{q}} + \alpha'_q t) } (1-x)^\beta ,
\end{eqnarray}
where the additional parameter, $\beta$, not shown in Table \ref{tab:parameters}, was allowed to vary in the interval, $\beta=1-7$
(a similar form was used in Ref.\cite{Hou:2019efy}).
%

Numerical results  are shown in Figure \ref{fig:formf} and Figure \ref{fig:PDF}. In Fig. \ref{fig:formf} we show our fit to the $n=1$ (FF) DS moments (upper panels), and to the CS, $n=2$ moments (lower panels). 
In Fig. \ref{fig:PDF}  we show results for the  $``-"$-symmetric components of the PDFs (upper panels), and for the $``+"$-antisymmetric distributions (lower panels). The key new results presented here are:  1) the identification of a potential disconnected contribution to the $``-"$ $x$-distributions, in addition to the valence-quark component, as illustrated in the upper panels of Fig.\ref{fig:PDF}; 2) the presence of a large disconnected contribution to the $``+"$ distributions. Note that a similar PDF model suggested in Ref.\cite{Hou:2019efy} showed a clear dominance of the DS over the CS at low $x$; however, the underlying GPD description presented here, offers more stringent, quantitative constraints rooted in the $x$ and $t$ correlated structure of these quantities. 

Note that although the magnitude of the DS term for $H_{u+d}^-$ is small, reflecting the constraint from the small values of the corresponding calculated FF \cite{Alexandrou:2025vto}, for $E^-_{u+d}$, based on the LQCD results, and lacking a constraint from the PDF in the forward limit, the contribution can be large: this could impact the interpretation of the flavor separated FF ratios originally displayed in Refs.\cite{Cates:2011pz,Arrington:2011kb,Qattan:2012zf,Qattan:2024pco}, which are also currently being measured at Jefferson Lab.

Another interesting aspect of the DS contributions to the FFs is in their role on the internal spatial distribution of quarks and antiquarks inside the proton. By Fourier transformation of the connected and disconnected GPDs, Eqs.\eqref{eq:GPD_minus} in 2D (\cite{Diehl:2002he} and references therein), we obtained the values of the average radii of the CS and DS per quark flavor (Table \ref{tab:radii}). Reflecting the shape of the form factors, the DS is located at smaller distances than the CS, with respect to the center of momentum of the proton. 
This picture contrasts with, for example, meson cloud models, where antiquark-containing configurations are predominantly found at larger radii.

\begin{table}[H]
    \centering
    \begin{tabular}{lcccc}
\hline
\hline
    {$\langle {\bf{{b}}}^{2}_{\mathrm{\perp}} \rangle ^{1/2}$} [fm] \quad \quad & $u$ \quad & $d$    \\
    \hline 
    \hline
      total & 0.596  & 0.721 \\
      connected & 0.601  &  0.734 \\
      disconnected & 0.290 &  0.289 \\
      \hline
      \hline
      \hline
          \end{tabular}
    \caption{Mean values of the partonic radii averaged over $x$ for connected, disconnected, and the sum of the connected and disconnected form factors.}
    \label{tab:radii}
\end{table}

%



Furthermore, can the disconnected diagrams behavior shed light on the  strangeness, $s \bar{s}$, asymmetry discussed {\it e.g.} in Refs.\cite{Diehl:2007uc,Maas:2017snj}?
A possible connection can be established with the baryon junctions picture at low $x$ \cite{Kharzeev:1996sq,Frenklakh:2023pwy}.
Heavy-Ion collision models, based on the valence quark picture, estimate the baryon to charge transport ratio to be between $0.5-0.7$ or even smaller, well below what is observed in experiment at RHIC \cite{starcollaboration2024trackingbaryonnumbernuclear}. 
This ratio is smaller than the naive expectation of unity due to excess production of anti-strange quarks over strange quarks at mid-rapidity. However, recent study showed that incorporating the baryon-junction picture via color reconnection increases the baryon to charge transport ratio to $0.99\pm 0.03$ compared to the case with the valence quark picture \cite{starcollaboration2024trackingbaryonnumbernuclear,ghb2-4y3s}. In the baryon junction picture, a junction can be transported to mid-rapidity and coupled to quarks coming from the $q\bar{q}$ pairs from the vacuum forming baryons resulting in the ratio close to unity. This gives us an intuition that the excess of anti-strangeness observed in experiments might originate from disconnected contributions at low $x$ \cite{Frenklakh:2023pwy}.  In a baryon junction picture including the $t$-dependence from GPDs, the observable for electron proton scattering processes becomes $ \propto \exp^{-\alpha \ln{\sqrt{\tau}}} \exp^{-\alpha y}$ where $\alpha$ is the Regge parameter and $y$ is the rapidity. This opens the way to a quantitative model, consistent with the intuition of Ref.\cite{Frenklakh:2023pwy}, which we will explore in future work.

In summary, using the concept of GPDs, we presented a study of the impact of disconnected contributions to the PDFs, constrained by the values of their first (form factors) and second Mellin moments. We find that disconnected contributions can impact the $``+"$, antisymmetric sea quark distributions, where their effect is not strictly dominant but is  more pronounced at low values of $x$. This initial study is a precursor to more questions on what generates the baryon-antibaryon asymmetry in QCD.

This work was completed under DOE grants DE-SC0016286 and DE-SC0024644. We acknowledge the fruitful discussions at the First Workshop on Baryon Dynamics from RHIC to EIC, at CFNS, Stony Brook (2024), and the Theory Group at Brookhaven National Lab. We thank  the members of the EXCLAIM collaboration for many discussions, in particular, Michael Engelhardt for pointing out the work in Ref.\cite{Syritsyn:2025fiu}.

\bibliography{references}

\end{document}